# QUASI-BOUND STATES AND CONTINUUM ABSORPTION BACKGROUND OF POLAR $Al_{0.5}Ga_{0.5}N$/GaN QUANTUM DOTS

D. Elmaghraoui[a], M. Triki[a], S. Jaziri[a,b], M. Leroux[c], J. Brault[c]

[a]*Laboratoire de physique de la matière condensé, Faculté des sciences de Tunis,*

*Campus universitaire 2092 El Manar, Tunisia*

[b]*Laboratoire de Physique des Matériaux, Faculté des Sciences de Bizerte 7021 Jarzouna, Tunisia*

[c]*Centre de Recherche sur l'Hetero-Epitaxie et ses Applications, Centre National de la Recherche Scientifique, Rue B. Gregory, 06560 Valbonne, France*

**ABSTRACT**

A theoretical interpretation of the photoluminescence excitation spectra of self-organized polar GaN/(Al,Ga)N quantum dots is proposed. A numerical method assuming a realistic shape of the dots and including the built-in electric field effects is developed to calculate their energy structure and hence their optical absorption. The electron and hole spectra show the existence of a set of quasi-bound states that does not originate from the wetting layer, and plays a crucial role in the observed absorption spectrum of the GaN/(Al,Ga)N dots. Transitions involving these quasi-bound states and wetting layer states give a sufficient explanation for the observed continuum absorption background. The properties of this absorption band, especially its extension, depend strongly on the dot's size. Our simulation provides a natural explanation of the experimental luminescence excitation spectra of ensembles of dots of different heights. Our theoretical model can be extended to cases where the confinement potentials are complicated by the presence of a stronger electric field.

## I. INTRODUCTION

Semiconductor heterostructures based on group-III nitrides are the base of numerous optoelectronic devices, due to their capacity for emitting light over a large visible and ultraviolet range [1-3]. A particularly interesting property of polar (*i.e.* grown along the polar axis of the wurtzite structure) III nitrides heterostructures is the strong redshift of the ground level transition when the nanostructures height increases [4]. This points towards the existence of a strong built-in electric field in the heterostructures, which causes a substantial quantum confined Stark effect [4,5]. Such intrinsic electric field originates from heterointerface discontinuities of the macroscopic bulk polarizations of the nitride systems.

Reducing the carrier's degrees of freedom in semiconductor-based optoelectronic systems has been beneficial to the development of semiconductor lasers. As a result, self assembled GaN quantum dots (QDs) are becoming a subject of increasing interest [6-11]. Such systems where excitons are confined in all three directions have been recently considered for UV-visible light emitters [7, 8]. The full discretisation of the energy levels inside QDs is of interest in order to improve the semiconductor laser characteristics.

The discrete levels scheme has been observed and discussed in materials such as the III-V arsenide system [12, 13] and later in II-VI QDs [14]. Near-field photoluminescence excitation (PLE) spectra of single QDs display a 2D-like continuum of states and a number of sharp lines between a large zero-absorption region and the 2D wetting layer (WL) edge [12,15]. Kammerer et al [15] have demonstrated that the observed continuum background in the up-converted photoluminescence signal of InAs/GaAs QDs is possibly related to the WL. More recently, another source of intrinsic broadening for the dot bound states was demonstrated theoretically [16]. Indeed, this broadening is related to the existence of transitions involving one bound state and one delocalized state near the dots.

Recently, a PLE experiment of single AlN/GaN dots, performed with a specially developed UV micro-photoluminescence excitation spectroscopy system, showed a resonant absorption into an excited state about 250 meV above the ground one, and a continuum of states below the WL ground state energy [9]. Je-Hyung Kim et al [10] have also studied the PLE spectra of $Al_{0.5}Ga_{0.5}N$/GaN QDs. The PLE spectra have also shown an absorption continuum present in all the samples studied, with variations in their edge energy, and for which there is still no reliable interpretation. The understanding of the fine structure of such systems becomes therefore of central importance. The prediction of the optical properties of GaN QDs is complex because it must include the effects of strain and piezoelectric fields, added to confinement effects. Many works have treated the optical structure of GaN QDs by several ways. For example, Andreev *et al.* have performed extensive calculations for truncated pyramidal AlN/GaN quantum dots using a **k.p** model in the envelope function approximation [17]. Other works, considering the truncated pyramidal shape of the polar nitride dots with a large diameter to height ratio, have approximated them to equivalent quantum wells (QWs) to calculate their ground state properties [11,18]. However, it should be noted that the main property of QDs is the carrier confinement in the three directions. Based on this, and since we are interested in determining the excited states of GaN/(Al,Ga)N QDs, we have chosen to describe their optical properties by a model that has a geometry respecting the three dimensional confinement and takes into account the built-in electric field. We will show later that the results obtained by this model are in good agreement with the experimental ones.

The aim of this work is thus to provide a theoretical interpretation of the PLE spectra of polar GaN/(Al,Ga)N QDs. Our model assumes a cylindrical symmetry of the dot which is reasonably realistic. Actually it can be applied to any arbitrary dot shape with cylindrical symmetry. It gives moreover the entire carrier spectrum at once. We find as expected electron and hole ground states spatially separated by the built in electric field with a strong reduction of their overlap integral. On another hand, crossed transitions, defined as a transition between one bound state and one delocalized state, and the transitions involving quasi-bound states provide a natural explanation of the continuous background measured in our luminescence excitation experiments. This continuous background exists with different extension in samples with different dot heights.

## II. EXPERIMENTS

The GaN/$Al_{0.5}Ga_{0.5}$N QDs samples studied were grown on sapphire by molecular beam epitaxy using $NH_3$ as N precursor. They consist of 3 QD planes separated by 30 nm $Al_{0.5}Ga_{0.5}$N barriers on a 1 μm thick relaxed $Al_{0.5}Ga_{0.5}$N template. An additional surface QD plane was deposited for atomic force microscopy purposes. Details on their growth and on their basic (ground state) luminescence properties can be found in [11]. The amount of deposited GaN was varied between samples, leading to different average dot heights. The buried dots have at truncated hexagonal pyramidal shape, as shown by transmission electron microscopy (TEM) studies [19]. The PLE spectra were measured using an Xe lamp filtered by a monochromator. On figure 1 are shown two luminescence spectra of our sample, with average dot height (including wetting layer) of 4 nm, as measured by TEM [19]. One of the PL spectrum was excited using 20 mW of the 244 nm line of a frequency doubled Ar laser. A broad band peaking near 3.2 eV corresponds to the QD ensemble, the transitions at 4-4.1 eV are attributed to recombinations in the wetting layer which, according to TEM, is 3 or 4 monolayers thick. The barrier luminescence is also observed at 4.6 eV. From the PL energy of the barriers, an Al content varying between 0.5 and 0.55 among the samples is deduced [20], in agreement with energy dispersive X ray analysis. The second PL spectrum shown in figure 1 is excited with 4.76 eV light from the Xe lamp used for the PLE experiments. The QD ensemble PL band is now peaking at 2.9 eV. The large difference between the QD peak energy between both spectra is due to an efficient screening of the Stark effect in the case of Ar laser excitation, while when using monochromated Xe light excitation (in the µW range), this screening is negligible [21]. Three excitation spectra are also shown in figure 1. They correspond to three different energies from the QD PL band, i.e. to three different QD sizes (labelled QD1, QD2, and QD3). The PLE spectra all show a resonance at 4.75-4.85 eV which corresponds to the (Al,Ga)N barrier absorption. The large difference in energy between barrier emission and barrier absorption is due to the fact that the emission involves localized excitons with holes from the barrier valence band maximum of $\Gamma_7$ symmetry, whose hole wave function is mainly $p_Z$-like (Z is parallel to the c axis) [20]. As such, transitions with electrons are almost forbidden for photons polarized along X or Y, which is the case of our experiments (either PL or PLE). On the other hand, PLE, i.e. absorption, monitors transitions strongly allowed for X and Y impinging photons, i.e. excitons involving the $\Gamma_9$ and the – unresolved- upper $\Gamma_7$ valence band maxima [20]. The PLE spectra also all show a broad absorption continuum, with a variation of in their edge energy. This continuum is strong enough so that no resonance corresponding to the wetting layer energy is observed. A resonance at 3.4 eV can be seen on the PLE spectrum corresponding to the lowest PL energy studied (QD1). The PLE spectra in figure 1 also show oscillations due to interferences originating from reflections at the air/(Al,Ga)N and (Al,Ga)N/sapphire interfaces. These interferences are not taken into account in our modeling. The aim of our work is to find the origin of these absorption spectra. To that end, we have considered QDs samples with different average heights characterized by the material and confinement parameters for electrons and holes given in table I.

### III. MODEL

The main features of polar wurzite GaN heterostructures are the large strain accumulated due to the large lattice parameter mismatches, and the strong built-in electric field, of the order of MV/cm, due to spontaneous and piezoelectric polarization discontinuities. To simplify the strain problem, we assume that the GaN is homogeneously biaxially compressed in the (0001) plane. The TEM study of AlN/GaN quantum dots by Arley et al [22] shows that this is a rather good approximation. This strong biaxial strain ($\varepsilon_{xx} \approx$ - 1.2 %) shifts the GaN excitonic gap to about 3.6 eV, which is the value we use in the following [23]. Note also that such a huge biaxial compression largely shifts downwards the lowest $\Gamma_7$ valence band state (corresponding to C excitons in GaN), while the upper $\Gamma_7$ one (corresponding to B excitons), stays near the $\Gamma_9$ one (corresponding to A excitons) with very similar wave functions (the separation is 11-12 meV, i.e. 2 thirds of the spin-orbit term). Since these two last valence band states are those optically active in normal incidence (k // c) experiments and since our experimental broadening is large (see below), we consider a single valence band in our calculations.

The built-in electric field has the same order of magnitude inside and outside the dot, but it is opposite in sign, antiparallel to the growth direction inside the dot for (0001) growth. The total potential V is the sum of two contributions: the confinement potential and the electric field effects. The calculation of the confinement potential depends strongly on the chosen shape. Transmission electronic microscopy [19, 22], have been used to characterize the morphology of the QDs. Buried polar GaN QDs in AlN or $Al_{0.5}Ga_{0.5}N$ matrices are truncated hexagonal pyramids. Since in the present work, we shall expand the dot wave functions with functions having cylindrical symmetry, the dot shape is approximated by truncated cones with basis radius $R_b$, top radius $R_t$, and height $h_2$ that are above a GaN wetting layer (WL) of thickness d . In the calculations, the dot is placed at the center of a large cylindrical box of radius R and height H. The domains that describe the QD structure and the wetting layer are given by the following expressions:

$$D(z,\rho) = D_{QD}(z,\rho) + D_{WL}(z,\rho)$$

$$D_{WL}(z,\rho) = \theta\left(\left(d + \frac{H}{2}\right) - z\right)\theta\left(z + \frac{H}{2}\right)$$

$$D_{QD}(z,\rho) = \theta\left(\frac{H}{2} + h_1 + d - z - \rho\tan(\alpha)\right)\theta\left(z - \left(d + \frac{H}{2}\right)\right)\theta\left(\frac{H}{2} + d + h_2 - z\right)$$

In these expressions $h_1$ and $\theta(x)$ denote respectively the cone's height, and the Heaviside function. For clarity the cross section of a GaN/(Al,Ga)N QD with its structural parameters is plotted in Fig.2. The carrier confining potential can be described in terms of an effective potential $V_e(V_h)$ for electron (hole) whose height is dictated by the conduction/valence band discontinuities. The origin of the potential is chosen to be the maximum of the valence band of the GaN bulk materials. Finally, the total potential for electron (hole) takes the following expressions:

$$V(\mathbf{r}_{e(h)}) = V_{e(h)}(1 - D(\rho,z)) \mp eF(\rho,z)z$$

Where e is the elementary charge. We consider a conduction to valence band offset ratio of $Q_e/Q_h = 70/30$ [24]. The energy band gap of the barrier material is calculated according to:

$$E_g^{(Al_xGa_{1-x}N)}(\rho,z) = (1-x(\rho,z)).E_g^{(GaN)} + x(\rho,z).E_g^{(AlN)} - C.x(\rho,z).(1-x(\rho,z))$$

$E_g^{(GaN)}$ and $E_g^{(AlN)}$ are the energy gap of GaN and AlN respectively. $x(\rho,z)$ is the aluminum barrier composition. The bowing parameter C is of the order of 0.9 eV for the $\Gamma_9$ gap of (Al,Ga)N [25]. In GaN/Al$_{0.5}$Ga$_{0.5}$N QDs grown by molecular beam epitaxy, it has been shown that there is a profile in the barrier's aluminum composition x, i.e. the distribution of the Al concentration is not uniform throughout the barrier region [18,26]: it is x = 0.5-0.55 in most of the barrier, but increases to x = 0.7-0.8 just above the QDs. In this work, we can take this distribution of concentrations into account, as shown in Fig.2.

In the total potential expression $F(\rho,z)$ is the electric field, whose spatial distribution is not uniform. Respecting the continuity of the electric displacement vector at the dot-barrier interfaces and the periodic boundary condition we can derive an expression of the spatial distribution of the electric field. We start with the conservation of the electric displacement vector across the heterostructure, which leads to:

$$\varepsilon_d F_d - \varepsilon_b F_b = P_b - P_d$$

Where $\varepsilon_d$ ($\varepsilon_b$) is the dot (barrier) dielectric constant, $F_d$ ($F_b$) is the electric field in the dot (barrier), and $P_b$ ($P_d$) is the total polarization in the barrier (dot) material. Assuming that the dielectric constant is uniform ($\varepsilon_d = \varepsilon_b = \varepsilon$) the previous equation is simplified and becomes:

$$F_d - F_b = \frac{P_b - P_d}{\varepsilon}$$

This expression corresponds to the total electric field discontinuity noted by $\Delta F$. The periodic boundary condition $\frac{F_b L_b(\rho)}{2} + F_d L_d(\rho) + F_b \left(\frac{L_b(\rho)}{2} - L_d(\rho)\right) = 0$, where $L_b(L_d)$ is the barrier (dot) width, leads to:

$$F(\rho,z) = \begin{cases} F_b = -\frac{\Delta F L_d(\rho)}{L_b(\rho)} & \text{if } z \in D(\rho,z) \\ F_d = \Delta F + F_b & \text{if } z \notin D(\rho,z) \end{cases}$$

The electric field values in each region of the system dot-barrier are obtained by replacing $L_d$ and $L_b$ by their appropriate values. In the case of an aluminum distribution of 0.75 above the dot we should take attention to the barrier electric field in this region. Using the equivalence of the dot's total potential at the edge of the large cylinder (z=0 and z=H) we can obtain the value of such electric field.

Within the usual effective-mass and the envelope-function approximation scheme, the electron and hole wave functions are described by the following Schrödinger equation:

$$\left(-\frac{\hbar^2}{2}\nabla \frac{1}{m_{e(h)}(z,\rho)}\nabla + V(\boldsymbol{r}_{e(h)})\right)\psi(\boldsymbol{r}_{e(h)}) = E^{e(h)}\psi(\boldsymbol{r}_{e(h)})$$

where $m_{e(h)}(z,\rho)$ is the bulk effective mass for electron (e) and hole (h) respectively. For a system with cylindrical symmetry, the total Hamiltonian commutes with the z projection of the total angular momentum *Lz*. Therefore, eigenstates solutions of the Schrödinger solution are also eigenstates of the z component of angular momentum

To solve the eigenvalue equation we used a wave functions expansion technique. The wave function $\psi(\boldsymbol{r}_{e(h)})$ is extended in terms of the complete Fourier-Bessel orthogonal set of solutions of a big cylinder that has the same axis of symmetry as the QD [27]:

$$\psi(\rho_{e(h)}, \theta_{e(h)}, z_{e(h)}) = \sum_{i,j} C_{n,i,j}^{e(h)} \varphi_{n,i,j}(\rho_{e(h)}, \theta_{e(h)}, z_{e(h)})$$

The orthogonal basis $\varphi_{n,i,j}(\rho_{e(h)}, \theta_{e(h)}, z_{e(h)})$ is given by:

$$\varphi_{n,i,j}(\rho_{e(h)}, \theta_{e(h)}, z_{e(h)}) = \sqrt{\frac{2}{V}} \frac{2}{|J_{n+1}(\lambda_i^n) - J_{n-1}(\lambda_i^n)|} J_n\left(\frac{\lambda_i^n \rho_{e(h)}}{R}\right) \sin\left(\frac{\pi j\, z_{e(h)}}{H}\right) e^{in\theta_{e(h)}}$$

In these expressions $J_n$ is the Bessel function of $n^{th}$ order, $n = 0, \pm 1, \pm 2, \ldots$, V is the volume of the big cylinder and $\lambda_i^n$ are obtained by solving the equation $J_n(\lambda_i^n) = 0$, where i and j $\in$ $Z^*$. By substituting the wave functions in the Schrödinger equation we obtain an eigenvalue equation. Energy levels and corresponding wave functions are then obtained by direct diagonalization of this equation. The states with different symmetries will be called according to the index $n$ or to the usual atomic physics nomenclature. In particular, the states are labeled S (n=0), P ($n = \pm 1$)..., so that the ground state and the first excited state are s-like and p-like states respectively. In this work we are limited to the S and P states.

To conclude our description, we simply summarize the used notation. After the diagonalization, our states are characterized by two quantum numbers ($\psi_{n,l}^{e(h)}$) : n indicates the symmetry of states (S, P) and l= 1,2,3 .... indicates its ordering. For example l = 1 is the first state l = 2 is the second state, etc… According to this, the energy levels are organized as follows: $E_{1S} < E_{1P} < E_{2S} < E_{2P} \ldots \ldots$

Let us comment on some aspects of our calculations; the choice of the dimension's matrix, the rayon and the height of the large cylinder box is not arbitrary but related to the convergence of the fundamental state. In this work we found that our system converge for $H = 600\text{Å}$, $R = 300\text{Å}$, and for a basis formed from 25 sines and 25 Bessel functions for each value of n.

## IV-DISCUSSIONS AND RESULTS
### A. Quasi-bound states

We present results on the three quantum dots (QD1, QD2 and QD3). The QDs sizes have been chosen close to experiments [19]. Thus we assume that the base radius of the dots vary like $R_b=h_2/0.34$. Detailed analyzed of valence and conduction bands, energy structures and states symmetry is developed on the highest dot in figure 1 (QD1) with homogenous aluminum barrier composition (x=0.5).

First of all, we start by analyzing the conduction and valence bands. Figs.3 (a) and (b) show respectively the variation of the electron conduction band and the heavy-hole valence-band edges along the (0001) direction (solid lines: profiles through the truncated cone center $\rho = 0$, dashed lines: band edges along the (0001) direction at $\rho = 70$Å, dotted lines: band edges along the (0001) direction through the wetting layer at $\rho = 200$Å. Figs 3(c) and (d) are the calculated 2D map of the conduction band and valence band edge for the GaN/AlGaN QD's respectively. The origin of energies is chosen to be the maximum of the valence band of the GaN materials. The blue regions in fig.3 (c) correspond to potential wells for electrons and the red regions in fig.3(d) correspond to potential wells for holes. These maps clearly illustrate that the built-in potential creates a well (darkest regions) at the bottom of the dot for holes and at the top of the dot for electrons. Wide potential wells in the plane of the WL are observed for electrons and holes with a magnitude of 2.9eV and -0.58eV respectively. Moreover, red and blue regions are also observed surrounding the dot. As shown in figure 3(d) the valence band map show a deep and wide well (along the growth direction) just below the dot base with a magnitude lying between -0.2 and -0.6 eV. A similar result is observed in the conduction band map (fig.3(c)): the potential well is just above the dot with a magnitude lying between 2.5eV and 3eV. Because the distribution of potential is somehow complex, we will hereafter analyze the energies levels of the electron and hole and their states symmetry.

Figs.4 (a) and (b) show the energy levels of electrons and holes respectively for QD1. Orbitals with S symmetry are plotted with black lines, those with P symmetry with red lines. These spectra show a discrete series of bound states followed by a quasi-continuum and a continuum of unbound levels, corresponding to the WL and the barrier, respectively. In fact, three energy regions are distinguished in each spectrum. For the hole spectrum the three regions are defined as follow: from -0.3 to -1eV the states are confined in the QD (bound states), beyond that the states are delocalized in the WL (quasi-continuum), above -1.15eV the states become delocalized in the 3D continuum of the (Al,Ga)N bulk barriers (continuum). For electron, the states lying between 2.3eV and 3.1eV are confined in the QD, beyond that they are delocalized in the WL, above 3.25eV the states become delocalized in the 3D continuum of the barriers. The set of states delocalized in the WL and barrier will be labeled continuum. This situation has already been demonstrated for the arsenide QDs [27, 28]. The difference introduced in our spectra in comparison with the arsenide ones is that the highest bound states are looking like a quasi-continuum. We note the existence of a new set of states for both electron and hole spectra that we will called *quasi bound states* (we will discuss after this denomination) with a very large ones for the hole. Compared to electron the hole potential separation between the QD and WL potentials is very important (figs.3 (a) and (b)) leading to a large set of quasi-bound states before shifting into the WL. In order to understand the origin of quasi-bound states we return to the valence and conduction band maps (figs.3 (c) and (d)). As mentioned before we have noted the existence of potential wells bellow and above the QDs for both holes and electrons. The quasi-bound states have energies higher than the QD potential but lower than the potential surrounding the dots so they can delocalize bellow and above the dot but they are not yet authorized to be in the WL. To illustrate this picture we present in fig.5 the probability density distribution for typical electron (hole) bound and quasi-bound states. Fig.5 (a) and (b) confirm that the bound hole wave function is localized at the bottom of the quantum dot, partly in the wetting layer, while the electron wave function is pushed up to the top. Fig.5 (c) and (d) show the probability densities for electron and hole quasi-bound states respectively. The electron quasi-bound state shown in fig.5(c) is localized partly at the bottom of the dot with an important extension above it. Compared to the hole first bound state the hole quasi-bound state presented in fig.5(d) is localized below the QD with some density in the dot center. It is worth noting, that the important localization of the hole below the QD is due to its important mass compared to that of the electron. As such, bound particles having the highest discrete energies are not completely confined in the QD; they can move along the growth direction in the region just above and below the dot so their denomination quasi-bound states.

### B. Continuum background and crossed transitions

The single-particle functions of electron and hole found in this way are used to calculate the exciton states within a configuration interaction scheme. For this purpose, we take the space spanned by the products of the single particle states, i.e $|\Psi^X_{mn,lk}\rangle = |\psi^e_{m,l}\rangle \otimes |\psi^h_{n,k}\rangle$.

However since the localizing potential including the electric field for carriers in a QD is much larger than the potential due to the Coulomb interaction, we may ignore the Coulomb correlations in the motion of the electron and hole and calculate the Coulomb interaction energy, $V^c$, for two fixed spatially distributed charged clouds (this latter can be treated as a perturbation). Then the energies of optical transitions read as:

$$E_{mn,lk} = E^e_{m,l} - E^h_{n,k} - V^{mn}_{lk}$$

where $V^{mn}_{lk}$ is the Coulomb interaction between the (m,l) electron state and the (n,k) hole state, that can be expressed as :

$$V^{mn}_{lk} = \langle \Psi^X_{mn,lk} | V^c | \Psi^X_{mn,lk} \rangle$$

$$= \int d\mathbf{r}_e \int d\mathbf{r}_h \int \psi^{e*}_{m,l}(\mathbf{r}_e) \psi^{h*}_{n,k}(\mathbf{r}_h) \frac{e^2}{\varepsilon|\mathbf{r}_e - \mathbf{r}_h|} \psi^e_{m,l}(\mathbf{r}_e) \psi^h_{n,k}(\mathbf{r}_h)$$

For the calculation of the coulomb interaction magnitude, we have chosen at first to calculate the coulomb interaction between states of S symmetry (n=0). Our evaluation shows that the interactions $V^{00}_{lk}$ are more important between bound states transitions lSe—kSh; it's about 28 meV, 34meV and 42meV for the fundamental transition of QD1, QD2 and QD3 respectively, and $V^{00}_{lk}$ becomes smaller and smaller gradually as the transitions involve extended states. The consideration made on the S transitions is also valid for other transitions symmetries. We can conclude that the coulomb coupling shifts downwards the fundamental transition but does not alter the exited transitions. Based on the forgoing, we have considered the coulomb interaction only for the fundamental transition.

The absorption and emission spectra exhibit expanded peaks. In most cases this broadening is modeled by Gaussian functions instead of the Dirac distributions as they adequately modeled this broadening. Consequently, for a given symmetry, interband absorption was calculated as follows:

$$\alpha^{int}_n(E) = \frac{g_n}{\sqrt{2\pi}\sigma_E} \sum_l e^{-\left(\frac{\left(E-(E^e_{n,l}+E^h_{n,k})\right)^2}{2\sigma^2_E}\right)} |\langle \psi^e_{n,l} | \psi^h_{n,k} \rangle|^2$$

Where $g_n = 1$ for n=0 and $g_n = 2$ for n=1. $\sigma_E$ defined the broadening of the absorption spectrum. In this work we have chosen $\sigma_E = 50 me$, this value is the experimental broadening as can be deduced for instance by the PL second order resonance observed in the PLE spectrum of figure 1. It should be noted that the overlap functions exert an optical selection rule: the only inter-band radiative transitions are those between states with the same symmetry. In addition to the energy positions and oscillator strengths of the radiative transitions, our simulations provide useful information on their nature. When evaluating the interband transition, one deals with different kinds of transitions: (i) the bound-to-bound (quasi-bound) transitions; (ii) the bound-to-continuum ''crossed'' transitions; (iii) the continuum-to-continuum transitions related to the WL and barrier delocalized states.

Fig.6 shows the calculated absorption spectrum (black line) of QD1 compared to the experimental one (magenta line), using the following parameters: $R_b$ =11nm, $h_2$=3.8nm, and d=1nm, assuming an homogenous aluminum barrier composition (x=0.5). In this figure vertical lines indicate the different transitions between bound states and continuum (the so called *crossed transition*) and the continuum to continuum transitions. Some of bound (quasi-bound) to bound (quasi-bound) transitions are indicated by arrows. As shown in fig.6 the calculated WL two-dimensional ground state transition is at 4.17eV, slightly higher in energy than the wetting layer PL transitions observed in figure 1. The absorption edge at 4.77eV corresponds to the barrier 3D transition.

Many bound and quasi-bound transitions exist for this dot below the first crossed transition, they are not apparent due to their low oscillator strength and the important broadening. Above the edge of the WLh-1Se transition, bound-to-bound (or quasi-bound) transitions are completely inserted in the background continuum. We can note that some of quasi-bound transitions are clearly visible. In particular the first observed quasi-bound transitions is between a hole quasi-bound state and the electron first quasi-bound state. The agreement between this transition and that observed experimentally is poor. We will see below that is due to the assumption of an homogenous aluminum barrier concentration (i.e. x = 0.5) while it has in reality a graded profile around the QD [19]. Since this will affect the band offset above the dot and consequently the quasi-bound states we will analyze this situation later. Let us now focus on the continuum region of the PLE spectra. The nature of the transitions in such a region is an interesting issue to be discussed. This continuum has already been analyzed in the arsenides system. In fact the continuous background corresponds to a multitude of transitions that is related to the existence of *crossed transitions* involving QD bound states and WL states delocalized near the dot [16]. In our system and as shown in fig.6, many kind of transitions participate to this continuum; bound to bound transitions, quasi-bound to bound transitions, bound to continuum transitions, and quasi-bound to continuum transitions. The intensity of the continuous PLE background is more influenced by the transitions involving quasi-bound states than the crossing transitions. This can be explained by two major raisons; first raison: the set of quasi-bound states is very wider compared to the continuum presented in the WL so we start with a huge number of transitions involving quasi-bound states, second: the quasi-bound states are weakly affected by the electric field and show a shape spreading along the growth direction (see fig.3 (c) and (d)) with lobes in the dot center, leading to an important overlap between electron and hole states i.e. a significant optical oscillator strength and thus intense absorption. To conclude this section, transitions involving quasi-bound states give an important contribution to the absorption spectrum. They reinforce the intensity of the PLE continuum, as they are less confined and less affected by the electric field.

In this step we shall focus on the lowest part of the spectrum, in particular the pronounced resonance around 3.4 eV. Following our previous discussion we have to calculate the spectrum by taking into account the Al enrichment above the dot (noted $x_{top}$, $x_{top}$=0.75). To obtain the same experimental fundamental energy, we have to change the dot height from $h_2$=3.8 nm to $h_2$=4 nm. The corresponding PLE spectrum is compared in fig.7 (a) to the spectrum calculated with an homogenous aluminum barrier composition ($x_{top}$=0.5) and to the experimental one. Fig.7 (b) and (c) show the distribution probability densities for the hole and electron states associated to the 3.4 eV resonance in the $x_{top}$=0.5 case and the $x_{top}$=0.75 case, respectively. We have also presented the electrons and holes densities separately in the inserted figures; top inserted figure: electron probability density, bottom inserted figure: hole probability density. As fig.7 (a) shows, the pronounced peak, that corresponds to a transition between the 1Pe state for the hole and the first quasi-bound state for the electron, is slightly blue shifted and is more pronounced compared to that presented in the first spectrum. We can note that in both cases ($x_{top}$=0.5 and $x_{top}$=0.75) this transition involves the electron first quasi-bound state. Fig.7 (c) shows the important spatial extension of this electron first quasi-bound state with lobes that extend near the bottom of the dot resulting with an important overlap with the 1P hole state, while the state presented in fig.7(b) is extended laterally without any lobes near the dot center leading to a lower overlap with the hole state. We can conclude that the electron state is responsible for the enhanced oscillator strength of the resonance at 3.4eV as it is more sensitive to the potential above the dot than the hole that is located at the bottom. Finally, taking into account the inhomogeneity of the barrier composition gives a better fit to the experimental PLE spectrum, confirming somehow the study of M. Korytov et al [19]. We will consider the Al enrichment above the dot for QD2 and QD3 in what follows.

To calculate the second PLE spectrum observed in fig.1 (green line) we have to consider a QD with a height of 3.4nm (QD2). The simulated result is presented by the black line in fig.8; the green line corresponds to the experimental one. Two major remarks can be drawn from this figure. The first is that the crossed transitions are closer to the fundamental transition, in agreement with the experimental result for such a dot; the WLh-1Se transition edge is at 3.45eV while the fundamental transition is at 2.95eV. The second is that fewer bound (quasi-bound) transitions are present in the region above the continuous background and there is no resolved peak in this region. This is can be explained by the size of the QD. This dot is smaller than the one considered previously, so it has less confined states; the states are rapidly shifted to the WL. We have less discrete states and consequently less resolved peaks. So we can conclude that the properties of the continuum, especially its extension, depend strongly on the dot's size i.e the confining potential. To confirm this we consider in fig.9 the simulation of the third absorption spectrum (blue line in fig.1). The dot (QD3) corresponding to this spectrum is smaller than the two others ($h_2$=2.8 nm). We note again that the continuum is closer to the fundamental transition with a fewer confined states below the first crossing transition. Only the fundamental and the first excited states are confined in the dot. The 1Sh-1Se and the 1Ph-1Pe transitions are more pronounced in this case, this due to the weakness of the electric field effect for small dots and leading to important oscillator strengths. With decreasing the QD height the confined energy levels are well separated with a reduction of their number which results in a reduction of the number of resolved peaks in the continuous background.

## V. CONCLUSION

To sum up, the bound-to-continuum *crossed transitions* give a satisfying explanation to the observed continuous absorption background. Transitions involving quasi-bound states also give an important contribution to the absorption spectrum. They enhance the intensity of the continuous background, as they are less confined and less affected by the electric field. The evolution of the QD's size affects the extension of the continuous absorption background. Our theoretical method can be convenient for future optical studies including systems with more complicated potential.

**Acknowledgements**

We are highly indebted to J. Renard and B. Gayral (INAC, Grenoble, France) for their indispensable hand in UV PLE experiments. We are also thankful for fruitful discussion with Dr. A. Vasanelli (Université Paris Diderot - Paris 7, France).

.

**Figure captions:**

**Fig. 1:** PL and PLE spectra of (Al,Ga)N/GaN quantum dots measured at 7 K. The gray line is a PL spectrum obtained using 20 mW of the 244 nm line of a frequency doubled Ar laser. Black line: PL spectrum excited with the filtered Xe lamp (4.76 eV). Magenta, green and blue lines are PLE spectra measured at 2.7, 2.95, and 3.17 eV respectively.

**Fig. 2:** The cross-sectional profile of the QD with its geometrical parameters and the distribution of the barrier's aluminum composition.

**Fig.3:** (a) and (b) conduction band and heavy-hole valence-band edges respectively along the (0001) direction. Solid lines: profiles through the truncated cone center (ρ=0). Dashed lines: band edges along the (0001) direction at $\rho = 70\text{Å}$. Dotted lines: band edges along the (0001) direction through the wetting layer at $\rho = 200\text{Å}$ (c) and (d) the calculated 2D map of the conduction band and valence band edges of the GaN/AlGaN QDs respectively.

**Fig. 4:** (a) Energy levels of the GaN quantum dots conduction band. (b) Energy levels of the GaN quantum dots valence band. The origin of energy is the maximum of the valence band of GaN. The nS and nP states are plotted in black and in red lines respectively.

**Fig.5:** Electrons and holes probability densities corresponding to: (a) the electron fundamental state (1Se), (b) the hole fundamental state (1Sh), (c) an electron quasi-bound state, (b) a hole quasi-bound state.

**Fig.6:** Comparison between calculated (black) and experimental (magenta) PLE spectra for the lower energy dot (QD1).

**Fig.7:** (a) Calculated Absorption spectrum of QD1. Black line: a not uniform barrier's Aluminum composition ($x_{top}$=0.75); Red line a uniform barrier composition ($x_{top}$=0. 5). (b)

and (c) the distribution probability density for the hole and electron states associated to the pronounced transition for $x_{top}$=0.5 and $x_{top}$=0.75 respectively.

**Fig.8:** Comparison between simulation and experimental results (QD2). Black lines: the calculation absorption spectrum; green line: the experimental absorption spectrum.

**Fig.9:** Comparison between simulation and experimental results (QD3). Black lines: the calculation absorption spectrum; blue line: the experimental absorption spectrum.

**Table.I:** Parameters of the QDs and substrate materials. $m_0$ is the free electron mass.

**Figures:**

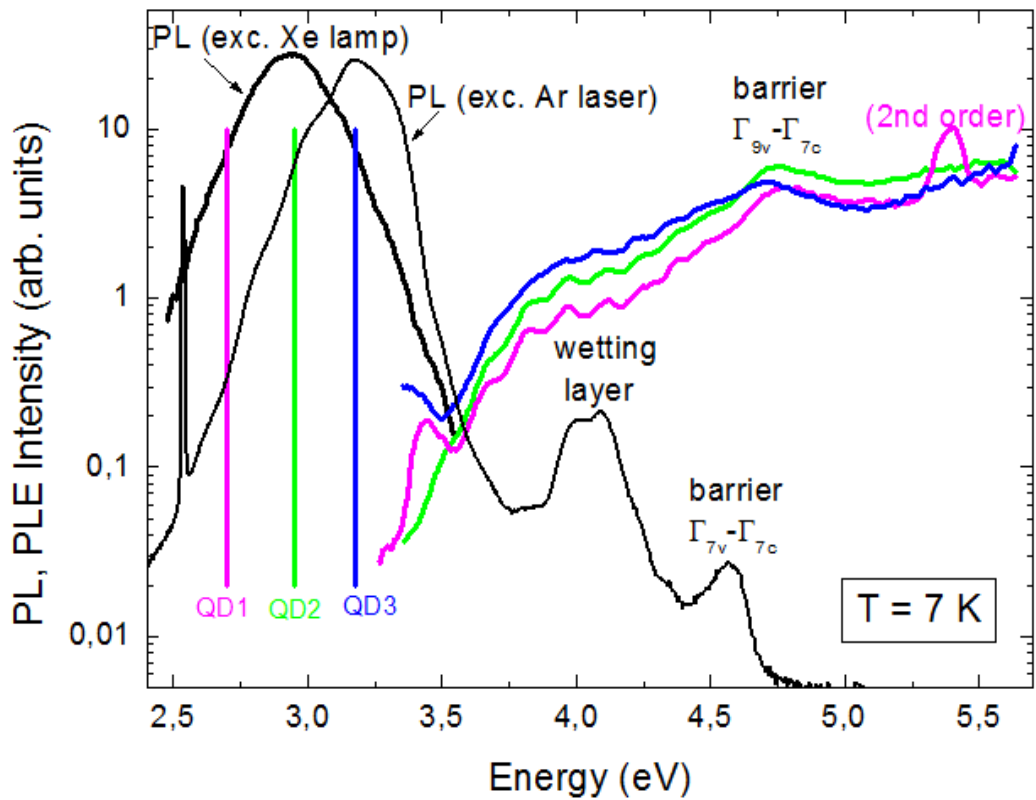

**Fig.1**

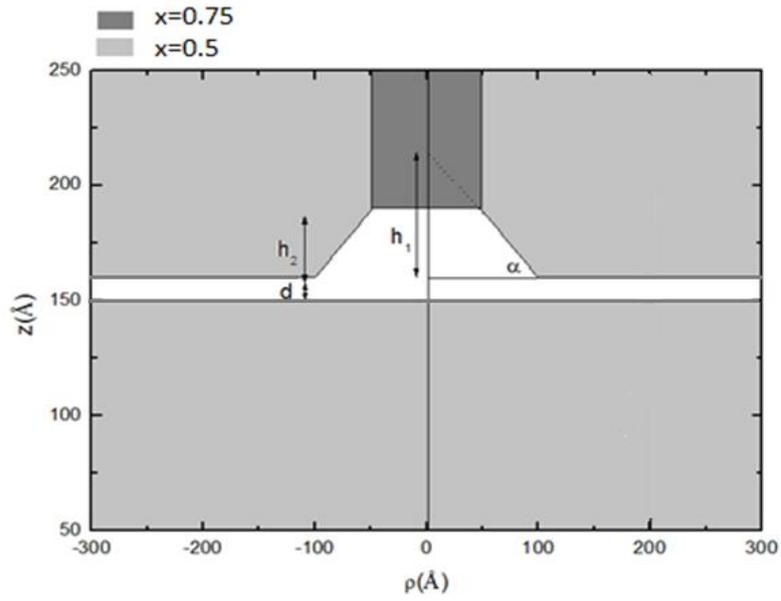

**Fig.2**

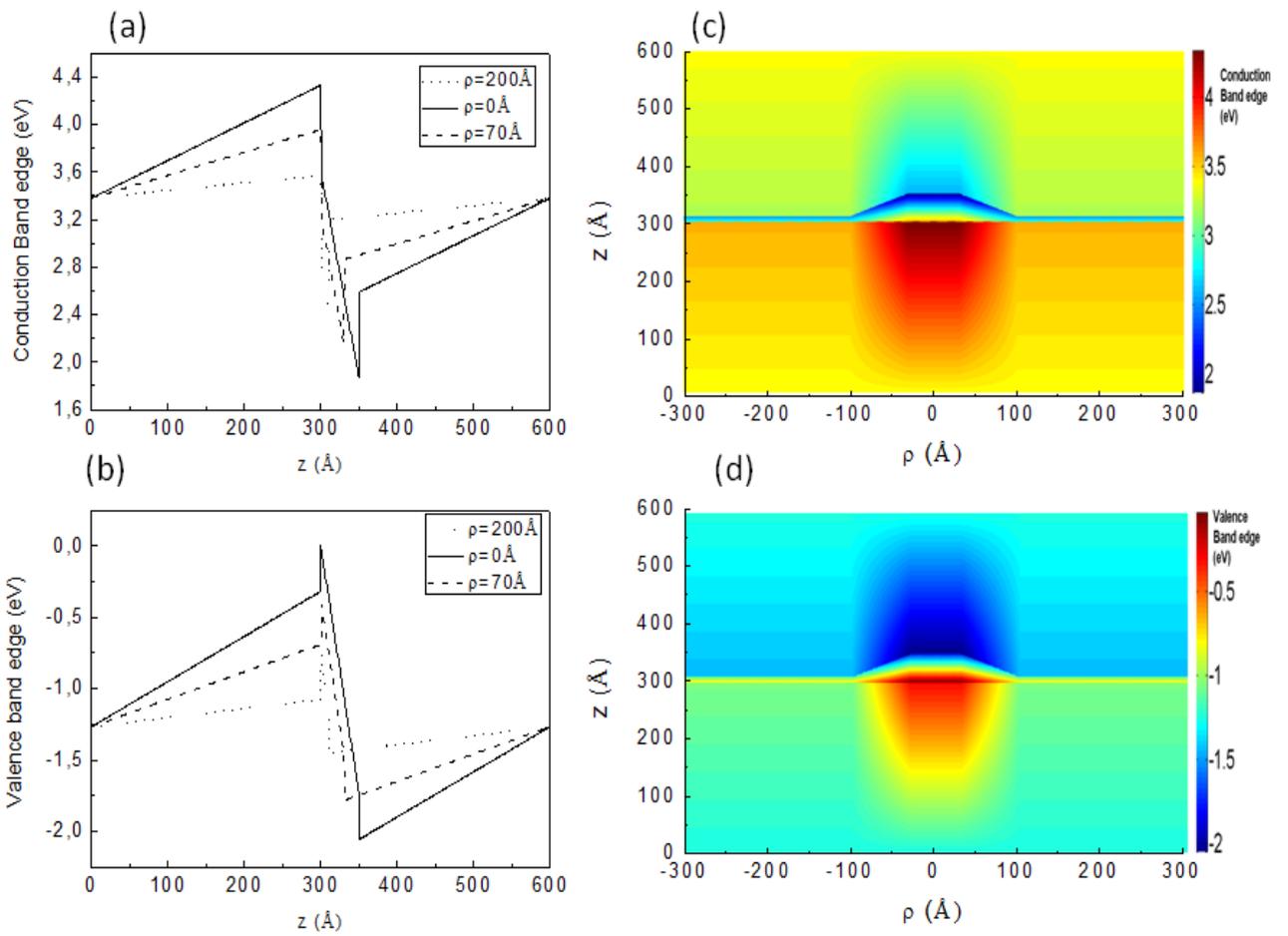

**Fig.3**

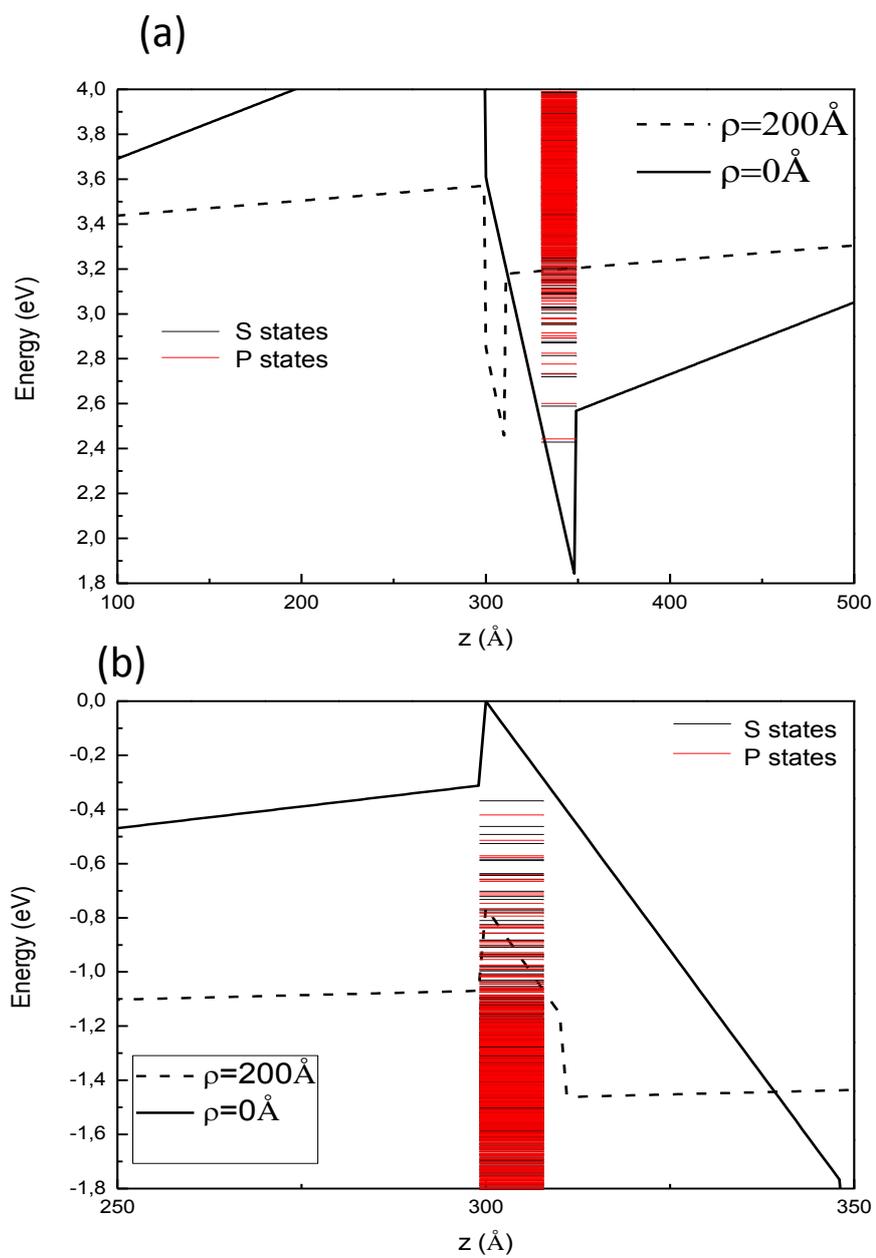

**Fig.4**

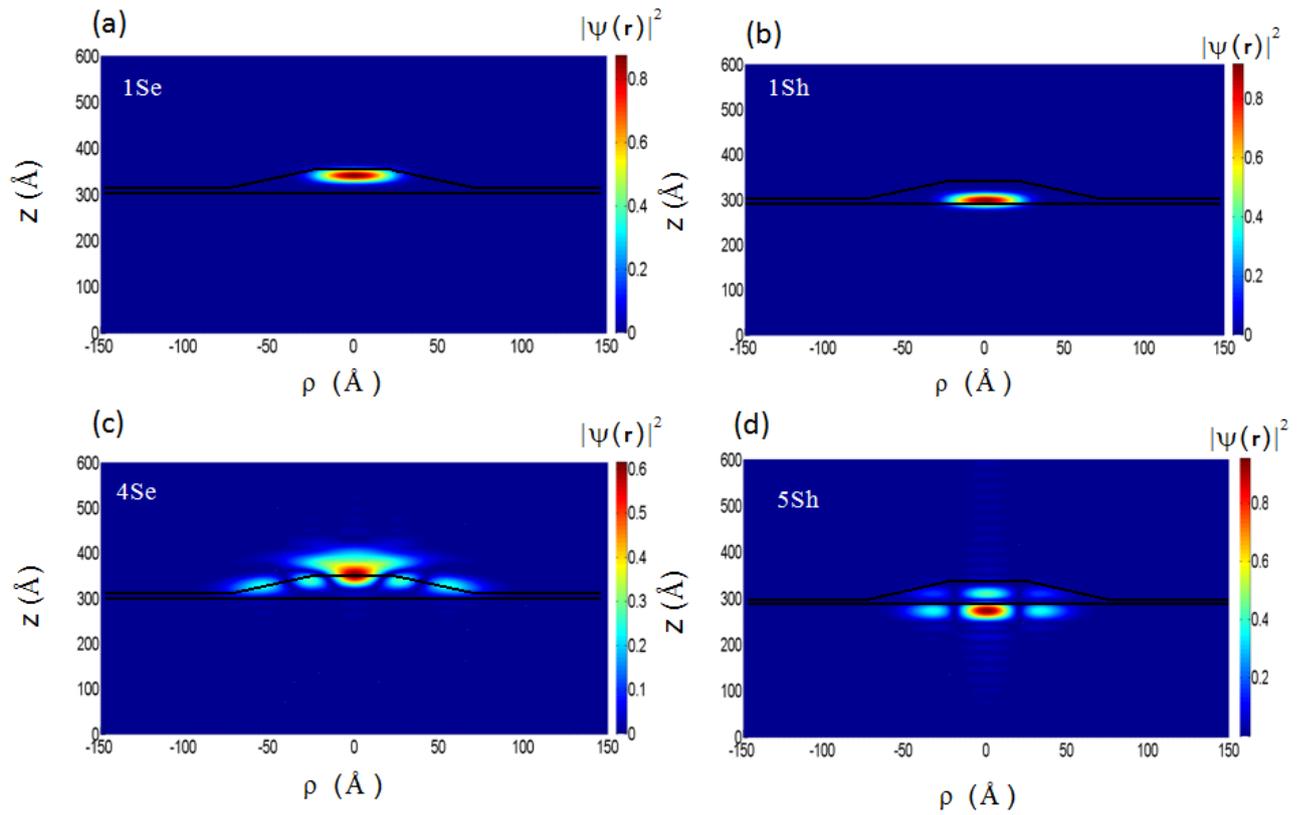

**Fig.5**

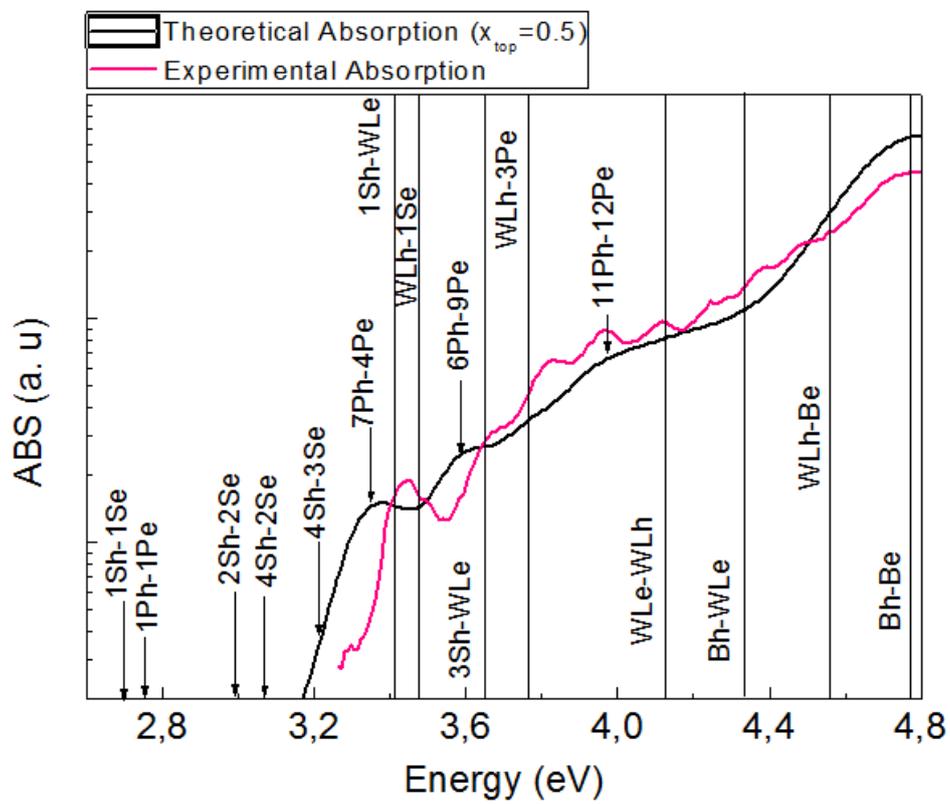

**Fig.6**

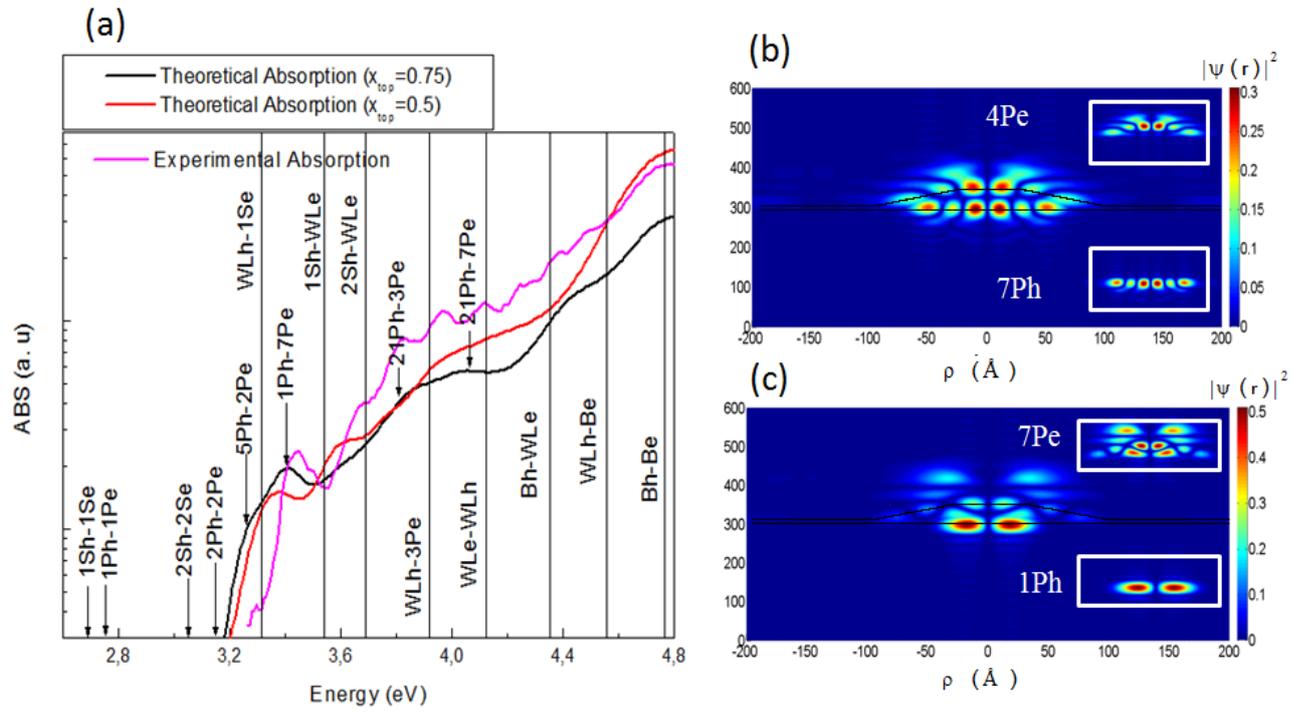

**Fig.7**

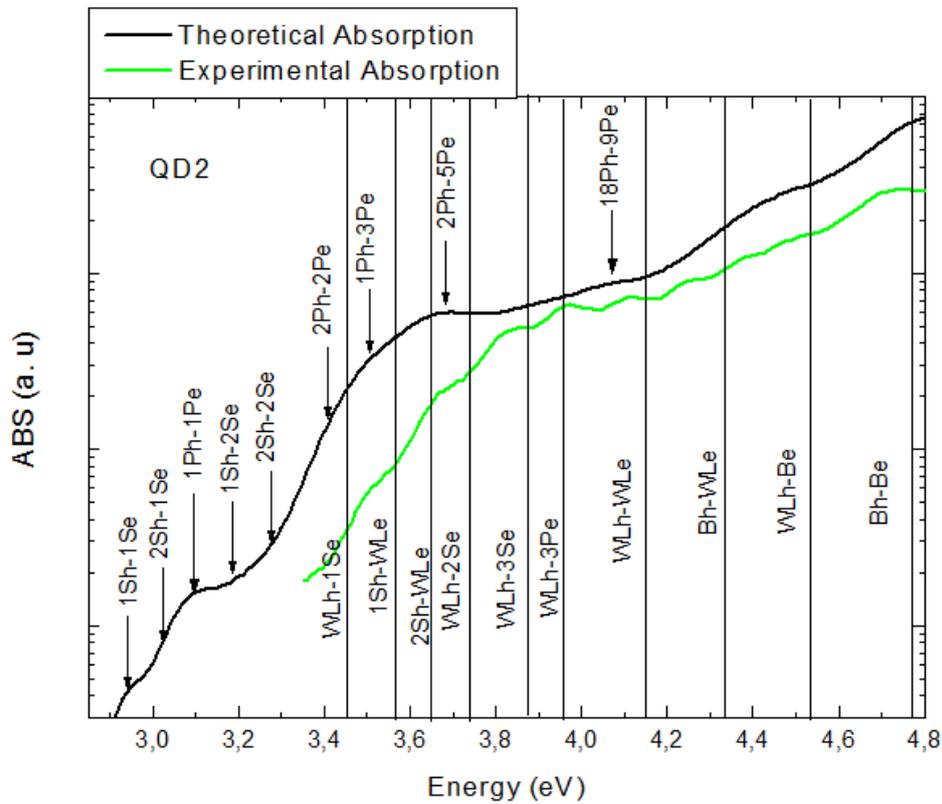

**Fig.8**

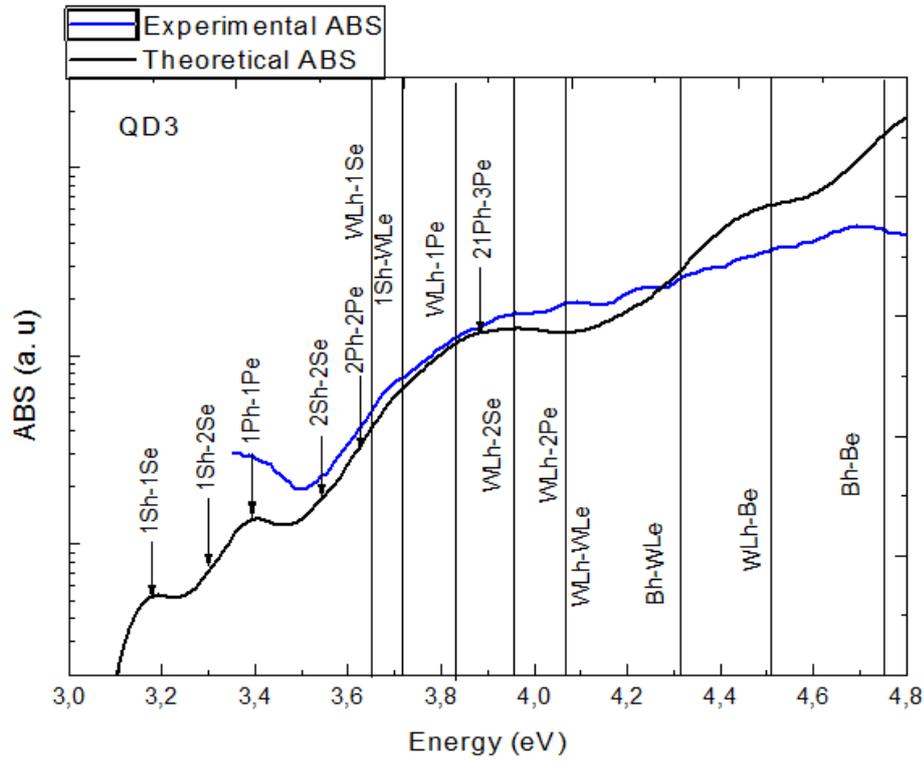

**Fig.9**

| $Eg^{GaN}$ | $Eg^{AlN}$ | $m_e^\perp$ ($m_0$) | $m_e^\parallel$ ($m_0$) | $m_h^\perp$ ($m_0$) | $m_h^\parallel$ ($m_0$) | $\Delta F$ | d | $h_2$(QD1) | $h_2$(QD2) | $h_2$(QD3) |
|---|---|---|---|---|---|---|---|---|---|---|
| 3.61eV | 6.28eV | 0.22 | 0.22 | 0.34 | 1.6 | 3.8MV/cm | 1nm | 4nm | 3.4nm | 2.8nm |

**Table I**